\documentclass[useAMS,usenatbib,usegraphicx]{mn2e}

\title[Electrostatic instabilities in the solar chromosphere]
{Electrostatic plasma instabilities driven by neutral gas flows in the solar chromosphere}
\author[G. Gogoberidze, Y. Voitenko, S. Poedts and J. De Keyser]
{G. Gogoberidze$^{1,2}$, Y. Voitenko$^{3}$, S. Poedts$^{4}$ and J. De Keyser$^{3}$\\
$^{1}$Dipartimento di Fisica, Universit\`a della Calabria, I-87036 Rende, Italy\\
$^{2}$Institute of Theoretical Physics; Ilia State University, Cholokashvili Ave 3/5,
Tbilisi, 0162, Georgia\\
$^{3}$ Solar-Terrestral Centre of Excellence, Belgian Institute for Space Aeronomy,
Ringlaan 3, 1180 Brussels, Belgium\\
$^{4}$ CPA/K.U.Leuven, Celestijnenlaan 200B, 3001 Leuven, Belgium}

\begin{document}

\date{Accepted . Received ; in original form }
\pagerange{}
\pubyear{2013}
\maketitle



\begin{abstract}
We investigate electrostatic plasma instabilities of Farley-Buneman (FB)
type driven by quasi-stationary neutral gas flows in the solar chromosphere.
The role of these instabilities in the chromosphere is clarified. We find
that the destabilizing ion thermal effect is highly reduced by the Coulomb
collisions and can be ignored for the chromospheric FB-type instabilities.
On the contrary, the destabilizing electron thermal effect is important and
causes a significant reduction of the neutral drag velocity triggering the
instability. The resulting threshold velocity is found as function of
chromospheric height. Our results indicate that the FB type instabilities
are still less efficient in the global
chromospheric heating than the Joule dissipation of the currents driving these instabilities. 
This conclusion does not exclude the possibility that
the FB type instabilities develop in the places where the
cross-field currents overcome the threshold value and contribute to the heating locally.
Typical length-scales of
plasma density fluctuations produced by these instabilities are determined
by the wavelengths of unstable modes, which are in the range $10-10^{2}$ cm
in the lower chromosphere, and $10^{2}-10^{3}$ cm in the upper chromosphere.
These results suggest that the decimetric radio waves undergoing scattering
(scintillations) by these plasma irregularities can serve as a tool for
remote probing of the solar chromosphere at different heights.
\end{abstract}


\section{Introduction}

Since it was discovered that the temperature in the solar chromosphere is
much higher than can be expected in radiative equilibrium, the mechanism of
chromospheric heating is one of the main puzzles in solar physics. The first
scenario for coronal and chromospheric heating was proposed by \citet{B46}
and \citet{S48}, who suggested that the atmosphere of the sun is heated by
acoustic waves generated in the turbulent convective zone. The theory of
wave generation by turbulence was developed by \citet{L52}. Extension of
this theory to the stratified environment of the solar atmosphere showed
that short-period acoustic waves are abundantly generated in the convective
zone \citep{S67}. The theory predicts that the peak of the acoustic power
spectrum is just below a period of one minute. Later numerical simulations
(e.g., \citet{CS92}) confirmed that the total power of the generated
acoustic waves is sufficient for chromospheric heating. But the measurements
of acoustic flux in the chromosphere have usually failed to find sufficient
energy. From the analysis of the Doppler shifts of UV lines, \citet{B78}
demonstrated that the energy flux of the acoustic waves with periods of 100
s or more is at least 2 orders of magnitude less than required for the
observed level of chromospheric heating. Similar results have been obtained
by \citet{MS81} from an analysis of the Doppler shifts of Ca II and Mg I
lines. Recent analysis of the data obtained by \emph{TRACE} \citep{FC05} has
shown that the observed intensity of high frequency (10-50 mHz) acoustic
waves was at least one order of magnitude lower than necessary for the
observed chromospheric heating. In addition, instead of steepening and
dissipation, the acoustic waves and pulses can form sausage solitons,
propagating undamped along magnetic flux tubes \citep{ZKK10}.

Problems with measurements of sufficient acoustic flux stimulated
development of alternative models of chromospheric heating. One of the
alternative scenarios \citep{P88,S99} implies that impulsive nano-flares
related to magnetic reconnection can be responsible for chromospheric
heating. The observations (e.g., \citet{A00}) do show numerous fast
brightenings in the sun but they are not sufficiently frequent to explain
the UV emission of the chromosphere. Another scenario for chromospheric
heating is resistive dissipation of electric currents \citep{RM84,G04}.
Recent analysis of three-dimensional vector currents observed in a sunspot
has shown that the observed currents are not sufficient to be responsible
for the observed amount of heating \citep{S07}.

Recently it has been supposed that a convective motion driven Farley-Buneman
instability \citep{F63,B63} (FBI) can significantly contribute to
chromospheric heating \citep{L00,F05,FPH08}. The FBI is known to be responsible for the formation of plasma
irregularities in the Earth's ionospheric E-region \citep{SN}. The interplay
of the background electric and magnetic fields at the altitudes where
electrons are strongly magnetized, produces currents that drive the
instability. In a similar way, if the electrons are strongly magnetized, the
drag of the ions by neutrals causes the instability. The simultaneously
observed electron heating was attributed to the parallel electric fields in
waves \citep{DM03,MD03}. 
\citet{GVPG09} extended analysis of the FBI in the solar chromosphere conditions by
taking into account the finite ion magnetization and Coulomb collisions. This study suggested 
that the FBI is not a dominant factor in the global 
chromospheric heating. However, local strong cross-field currents can drive
FBI producing small-scale ($0.1-3$ m) density irregularities and contributing to the chromospheric heating 
locally. \citet{PW13} accounted for the flow inhomogeneity (flow
shears) and found an electromagnetic MHD-like instability generated at
larger scales. These irregularities can cause scintillations of radio waves
at similar lengths scales and provide a tool for chromospheric remote sensing. 
It has to be noted that \citet{GVPG09} did not take into account effects of the electron heating related to the presence of parallel electric fields in the waves. As showed theoretically by \citet{DM03} and confirmed by recent particle in cell simulations \citep{OD13}, this effect can significantly increase the electron heating. Importance of this mechanism for the solar chromosphere requires an additional analysis and is beyond the scope of this paper.

It is also known that electron and ion thermal effects can strongly affect
small-scale E-region instabilities. The electron thermal effects lead to a
considerable modification of the FBI (mainly by the electron Pedersen
conductivity via perturbed Joule heating), and \citet{DS95} have given the
modified FB instability a new name: electron Pedersen conductivity
instability (EPCI). Later on, this instability was studied in more detail by %
\citet{DS97} and \citet{R98}. The ion thermal effects also modify FBI
significantly and make it possible in a wider altitude range as compared to
the predictions of adiabatic and isothermal FBI models \citep{DO04}.

Here we study small-scale electrostatic instabilities of the Farley-Buneman
type in the partially ionized plasma of the solar chromosphere taking into
consideration ion and electron thermal effects, electron and ion viscosity,
and Coulomb collisions. As it has been demonstrated by \citet{GVPG09},
contrary to the ionospheric case, the Coulomb collisions of electrons and
ions can not be ignored in the chromosphere because of the relatively high
degree of ionization ($10^{-2}-10^{-4}$). In the present paper we find
another difference with the ionosphere: the destabilizing influence of ion
thermal effects is highly reduced in the chromosphere by Coulomb collisions
and can be neglected. But electron thermal effects appeared to be important,
especially in the middle and upper chromosphere, where they reduce the
threshold value of the relative electron/ion velocity (current velocity). We
determine various characteristic length scales as well as the value of the
threshold relative velocity of electrons and ions necessary to trigger the
electrostatic instability as a function of chromospheric height in the
framework of the semi-empirical chromospheric model SRPM 306 \citep{FBH07}.
We confirm our previous conclusion that FB type electrostatic instabilities
cannot be responsible for the chromospheric heating at global length scales.
However, such instabilities can be generated locally in the places of
sufficiently strong currents and can create small-scale plasma
irregularities.

The paper is organized as follows. The general formalism is presented in
Sec. 2. The FBI and the ion thermal instability are studied in Sec. 3. The
electron thermal instability is discussed in Sec. 4. Different length scales
of the chromosphere important for the development of electrostatic
instabilities are studied in Sec. 5. Conclusions are given in Sec. 6.

\section{General formalism}

We use a standard modal analysis for linear perturbations in partially
ionized plasmas with neutral flows taking into account Coulomb collisions,
ion and electron viscosity, and thermal effects. The dynamics of electrons,
one species of singly charged ions and neutral hydrogen in the solar
chromosphere for imposed electric ($\mathbf{E}$) and magnetic ($\mathbf{B}$)
fields is governed by the continuity, Euler and heat transfer equations
\begin{equation}
\frac{\mathrm{d}_{\alpha }n_{\alpha }}{\mathrm{d}t}+n_{\alpha }\nabla \cdot
\mathbf{V}_{\alpha }=0,  \label{eq:1}
\end{equation}%
\begin{eqnarray}
m_{e}\frac{\mathrm{d}_{e}\mathbf{V}_{e}}{\mathrm{d}t} =-e\left( \mathbf{E}+%
\frac{\mathbf{V}_{e}\times \mathbf{B}}{c}\right) -\frac{\nabla (n_{e}%
\mathcal{K}T_{e})}{n_{e}}-  \nonumber \\
m_{e}\nu _{ei}(\mathbf{V}_{e}-\mathbf{V}_{i})-m_{e}\nu _{en}(\mathbf{V}_{e}-\mathbf{V}_{n})+m_{e}{\eta }_{e}\nabla ^{2}\mathbf{V}_{e},  \label{eq:2}
\end{eqnarray}%
\begin{eqnarray}
m_{i}\frac{\mathrm{d}_{i}\mathbf{V}_{i}}{\mathrm{d}t} =e\left( \mathbf{E}+%
\frac{\mathbf{V}_{i}\times \mathbf{B}}{c}\right) -\frac{\nabla (n_{i}%
\mathcal{K}T_{i})}{n_{i}}-  \nonumber \\
m_{e}\nu _{ei}(\mathbf{V}_{i}-\mathbf{V}_{e})-\mu _{ni}m_{i}\nu _{in}(%
\mathbf{V}_{i}-\mathbf{V}_{n})+m_{i}{\eta }_{i}\nabla ^{2}\mathbf{V}_{i},
\label{eq:2a}
\end{eqnarray}%
\begin{eqnarray}
n_{e}^{2/3}\frac{\mathrm{d_{e}}(T_{e}n_{e}^{-2/3})}{\mathrm{d}t} =\frac{2}{%
3}\varepsilon _{e}\mu _{ne}m_{e}\nu _{en}(\mathbf{V}_{e}-\mathbf{V}_{n})^{2}-
\nonumber \\
2\mu _{en}\nu _{en}\left( 1+\rho _{en}\right) (T_{e}-T_{n})+\frac{2}{3}\mu_{ie} m_{e}\nu _{ei}(\mathbf{V}_{e}-\mathbf{V}_{i})^{2}-  \nonumber \\
2\mu _{ei}\nu _{ei}(T_{e}-T_{i})+\frac{\chi _{e}}{n_{e}}\nabla ^{2}T_{e},
\label{eq:3a}
\end{eqnarray}%
\begin{eqnarray}
n_{i}^{2/3}\frac{\mathrm{d_{i}}(T_{i}n_{i}^{-2/3})}{\mathrm{d}t} =\frac{2}{3}\varepsilon _{i}\mu _{ni}m_{i}\nu _{in}(\mathbf{V}_{i}-\mathbf{V}%
_{n})^{2}-2\mu _{ni}\nu _{in}(T_{i}-T_{n})+  \nonumber \\
\frac{2}{3}\mu _{ei}m_{i}\nu _{ei}(\mathbf{V}_{e}-\mathbf{V}_{i})^{2}-2\mu
_{ei}\nu _{ei}(T_{i}-T_{e})+\frac{\chi _{i}}{n_{i}}\nabla ^{2}T_{i},
\label{eq:3}
\end{eqnarray}%
Here $\alpha =e,i$ denotes electrons or ions; $n$ denotes neutrals; $%
n_{\alpha }$ is the number density, $\mathbf{V}_{\alpha }$ is the averaged
drift velocity; $m_{\alpha }$ is the mass; $T_{\alpha }$ is the temperature;
$\nu _{\alpha \beta }$ is the elastic collision frequency; ${\eta }_{\alpha }
$ is the kinematic viscosity; $\chi _{\alpha }$ is the thermal conductivity;
$\mu _{\alpha \beta }=m_{\alpha }/(m_{\alpha }+m_{\beta })$ is the
mass-reducing factor, such that $2\mu _{\alpha \beta }$ is the energy
fraction lost by a particle of $\alpha $ species during one elastic
collision with a particle of $\beta $ species; $c$ is the speed of light; $%
\mathcal{K}$ is the Boltzmann constant, and $\mathrm{d_{\alpha }}/\mathrm{d}t
$ denotes the convective derivative. $\varepsilon _{e,i}$ are dimensionless
parameters which will be discussed below. The relative efficiency of
inelastic/elastic collisions in the electron thermal balance is $\rho _{en}=%
\bar{\nu}_{en}/\left( 3\mu _{en}\nu _{en}\right) $, where $\bar{\nu}_{en}$
is the inelastic $e-n$ collisional frequency.

Eqs. (\ref{eq:1})-(\ref{eq:3}) are similar to so-called '5-moment' transport
equations \citep{SN} which are often used when studying instabilities in the
E-region of the Earth's ionosphere. The principal difference between the
5-moment approach and our study is that, as it was mentioned in the
introduction, the ionization degree in the chromosphere is much higher than
in the E-region and consequently Coulomb collisions are not ignored in the
set of Eqs. (\ref{eq:1})-(\ref{eq:3}). We account for inelastic $e-n$
collisions \citep{R98} in the electron energy balance (\ref{eq:3a}) (term
proportional to $\rho _{en}$). We will come back to this last issue in the
discussion section.

The right hand side of Eqs. (\ref{eq:3a}) and (\ref{eq:3}) describe the
balance between frictional heating (two positive terms) and collisional
cooling (two negative terms). Without these effects the temperature
fluctuations would be adiabatic ($T_{\alpha }\sim n_{\alpha }^{\gamma -1}$
with $\gamma =5/3$). In the case of elastic collisions we have $\mu
_{ei}=m_{e}/(m_{e}+m_{i})\approx m_{e}/m_{i}$, and $\mu _{ni}\approx
m_{p}/(m_{p}+m_{i})$.

In the upper chromosphere the charged particles are mainly protons (and
therefore $\mu _{ni}=1/2$), whereas at lower attitudes heavy ions dominate
the positive charge. Because of this reason we do not specify the type of
ions and the obtained results will be suitable for studying both upper and
lower chromosphere. This circumstance leads to another distinctions from the
similar ionospheric analysis. Namely, for lower chromosphere we have $\mu
_{ni}\approx m_{p}/m_{i}$ and the influence of the ion-neutral friction on
ion dynamics is reduced by factor $m_{p}/m_{i}$ in comparison to the case of
equal ion/neutral masses.

The frictional heating terms in Eqs. (\ref{eq:3a})-(\ref{eq:3}) include
additional factors $\varepsilon _{e,i}$ that account for possible effects of
enhanced wave heating. Ionospheric observations shows that the typical value
of $\varepsilon _{e}$ varies between $10$ and $30$ in the middle ionosphere %
\citep{R98,DM03}, whereas no ion heating is usually observed (i.e., $%
\varepsilon _{i}=1$ in the ionosphere). Chromospheric factors driving waves
unstable appeared to be quite different from the ionospheric ones, and the
enhanced ion heating by the waves may occur in the chromosphere as well as
the enhanced electron heating. We would like to account for this possibility
by putting $\varepsilon _{i}\neq 1$, and for simplicity we will use the
single heating parameter $\varepsilon =\varepsilon _{e}=\varepsilon _{i}$.

For collision frequencies we use the following expressions \citep{B65}:
\begin{equation}
\nu _{ei}=\frac{4(2\pi )^{1/2}e^{4}n_{e}\Lambda }{3m_{e}^{1/2}(\mathcal{K}%
T_{e})^{3/2}},  \label{eq:c1}
\end{equation}%
\begin{equation}
\nu _{en}=\sigma _{en}n_{n}\sqrt{\frac{\mathcal{K}T_{e}}{m_{e}}},
\label{eq:c2}
\end{equation}%
\begin{equation}
\nu _{in}=\nu _{pn}=\sigma _{in}n_{n}\sqrt{\frac{\mathcal{K}T_{p}}{m_{p}}},
\label{eq:c3}
\end{equation}%
where $\Lambda $ is the Coulomb logarithm. From the former equation we see
that regardless of the mass of dominant ion species, $\nu _{ei}=\nu _{ep}$
for singly charged ions.

For the electron-neutral and ion-neutral collisions we assume a
simple model with constant cross-sections $\sigma _{en}=3.0\times 10^{-15}~%
\mathrm{cm^{2}}$ \citep{BK71} and $\sigma _{in}=2.8\times 10^{-14}~\mathrm{%
cm^{2}}$ \citep{KS99} that are typical for the middle chromosphere with
particles energies $\sim 0.5-1.0$ eV. In principle, $\sigma _{en}$ and $%
\sigma _{in}$ are not constant but depend on the particles energies. For
example, the neutral atom polarisation results in the $\sigma _{sn}\sim
1/V_{s}$ dependence making the collisional frequency indepent of the
particle energy. With this model our results would even more emphasise the
effects of Coulomb collisions on FBI in the upper chromosphere. However,
because of the other kinds of collisions with neutrals, the atom
polarisation model underestimates the electron and ion collisions with the
neutrals. Since these other kinds of collisions with neutrals are not well
studied in the chromospheric conditions, we use the model with constant
cross-sections, which artificially enhances $\nu _{en}$ and $\nu_{in}$ at larger heights.

Estimation of inelastic electron-hydrogen collisional frequency $\bar{\nu}%
_{en}$ is rather involved and sensitive to the electron temperature and
velocity distribution in the super-thermal tail. Taking into account two
main excitation levels of hydrogen atoms and using formulae given by %
\citet{J72}, we estimate that $\rho _{en}$ vary from $0.1$ in the lower
chromosphere to about 1 in the upper chromosphere. We will keep $\rho _{en}$
in derivations, but will not analyze its influence separately (see
Discussion).

We assume that the system is penetrated by a uniform magnetic field $\mathbf{%
B}$ and that neutrals have background velocity $\mathbf{V}_{n}\perp \mathbf{B%
}$. Then equation (\ref{eq:2})-(\ref{eq:2a}) give for the background flow of
electrons and ions
\begin{equation}
\kappa \mathbf{V}_{i}\times \mathbf{b}-(\alpha N+\mu _{ni})\mathbf{V}%
_{i}+\alpha N\mathbf{V}_{e}+\mu _{ni}\mathbf{V}_{n}=0,  \label{eq:4a}
\end{equation}%
\begin{equation}
-\kappa \mathbf{V}_{e}\times \mathbf{b}-\alpha (N+1)\mathbf{V}_{e}+\alpha N%
\mathbf{V}_{i}+\alpha \mathbf{V}_{n}=0,  \label{eq:4b}
\end{equation}%
Here $\kappa =\omega _{cp}/\nu _{pn}$ is the proton magnetization, $\mathbf{b%
}=\mathbf{B}/B$ is the unit vector along the mean magnetic field direction,
$\psi =\nu _{en}\nu _{in}/\omega _{cp}\omega _{ce}$, $\omega _{c\alpha
}\equiv eB/m_{\alpha }c$ is the cyclotron frequency, $\alpha =\psi \kappa
^{2}=m_{e}\nu _{en}/m_{p}\nu _{pn}\approx 2.6\times 10^{-3}$, and $N=\nu
_{ei}/\nu _{en}$ is the ratio of the Coulomb and electron-neutral collision
frequencies.

Multiplying equations (\ref{eq:4a})-(\ref{eq:4b}) by $\times \mathbf{b}$ and
excluding $\mathbf{V}_{i}\times \mathbf{b}$ and $\mathbf{V}_{e}\times
\mathbf{b}$ we get
\begin{eqnarray}
\left[ \frac{\kappa^2}{\mu_1}(N+1) + \alpha N +\mu_{ni} \right] \mathbf{V}%
_{i} - N\left[ \frac{\kappa^2}{\mu_1} + \alpha \right] \mathbf{V}_{e}=
\nonumber \\
\kappa \mathbf{V}_{n} \times \mathbf{b} +\mu_{ni} \mathbf{V}_{n},
\label{eq:4c}
\end{eqnarray}
\begin{eqnarray}
N\left[ \frac{\kappa^2}{\mu_1} + \alpha \right] \mathbf{V}_{i}- \left[ \frac{%
\kappa^2(\alpha N+\mu_{ni})}{\alpha \mu_1} + \alpha(1+N) \right] \mathbf{V}%
_{e}=  \nonumber \\
\kappa \mathbf{V}_{n} \times \mathbf{b} -\alpha \mathbf{V}_{n}.
\label{eq:4d}
\end{eqnarray}
Here $\mu_1=\alpha N+\mu_{ni}(1+N)\approx \mu_{ni}(1+N)$.

Using equations (\ref{eq:4c})-(\ref{eq:4d}) one can readily derive
expressions for $\mathbf{V}_{i}$ and $\mathbf{V}_{i}$, but exact relations
are too complicated. The dependence of the proton magnetization $\kappa $
and $N$ on height based on the semi-empirical chromospheric model SRMP 306 %
\citep{FBH07} is shown in Figs. \ref{fig2} and \ref{fig1}, respectively.
Detailed analysis of these data is presented in the next section. Here we
note that, as it can be seen from Fig. \ref{fig1}, for all chromospheric
heights $\alpha N\ll 1$. Also, from the data shown in Figs. \ref{fig2} and \ref{fig1} one can find that $\kappa ^{2}/\mu _{1}\gg \alpha $
in the chromosphere except for very low altitudes $h<600$ km. In this paper we are mostly interested in higher altitudes where the
Coulomb collisional effects are important for FBI. In the limit $\alpha
N\ll 1$ and $\alpha \mu _{1}/\kappa ^{2}\ll 1$ we obtain
the ion and electron background velocities
\begin{equation}
\mathbf{V}_{i}\approx \frac{\mu _{ni}}{\kappa ^{2}+\mu _{ni}^{2}}\left[ \mu
_{ni}\mathbf{V}_{n}+\kappa \mathbf{V}_{n}\times \mathbf{b}\right] ,
\label{eq:4e}
\end{equation}%
\begin{equation}
\mathbf{V}_{e}\approx \alpha \mu _{ni}\frac{N\kappa ^{2}+\alpha \mu
_{ni}(1+N)}{\kappa ^{2}(\kappa ^{2}+\mu _{ni}^{2})}\mathbf{V}_{n}-\alpha
\frac{\kappa ^{2}+\mu _{ni}^{2}(1+N)}{\kappa (\kappa ^{2}+\mu _{ni}^{2})}%
\mathbf{V}_{n}\times \mathbf{b}.  \label{eq:4f}
\end{equation}

In the considered limit, $\alpha \ll 1$ and $\alpha N\ll 1$, the current
velocity $\mathbf{U}_{0}=\mathbf{V}_{i}-\mathbf{V}_{e}\approx \mathbf{V}_{i}$.

On the background given by Eqs. (\ref{eq:4e})-(\ref{eq:4f}) and
corresponding solutions for ion and electron temperatures in the subsequent
sections we consider different linear electrostatic perturbations
propagating in the plane perpendicular to the background magnetic field. To
simplify further analysis, we make two assumptions which are standard in the
study of low frequency perturbations. Firstly, we assume quasi-neutrality $%
(n_{e}\approx n_{i}=n)$. This condition is valid when characteristic
frequency of perturbations is much less than ion plasma frequency. Secondly,
we treat electrons as inertialess. The latter assumption implies that the
characteristic time scale of the perturbations is much greater than electron
cyclotron and plasma time scales. Both ion-thermal and current driven
instabilities occur at ion-neutral collision time scale, which for typical
chromospheric parameters is much greater than all characteristic time scales
mentioned above.

In the analysis below we ignore perturbations of the neutral component. Such
a treatment is valid in weakly ionized plasma for relatively high frequency
perturbations. Comparing inertial and ion drag terms in the Euler equation
for neutrals, we obtain that perturbations of the neutral component can be
safely neglected if
\begin{equation}
\omega \gg \frac{n_e}{n_n} \nu_{in}.  \label{eq:7a}
\end{equation}

And finally, as is usually done in the E-layer research, we consider the ion
and electron temperature perturbations separately. Due to the relatively
high electron concentration in the chromosphere we do not ignore Coulomb
collisions. In the general case, perturbations of the ion temperature can
cause perturbations of the electron temperature. But due to the large
ions/electron mass ratio, Coulomb collisions are inefficient in the heat
transfer between electrons and ions. Mathematically this is manifested by
the $\mu _{ei}\sim m_{e}/m_{i}$ multiplier in the last but one term of
equation (\ref{eq:3}). Comparing this term with the left hand side of Eq. (%
\ref{eq:3}) shows that the thermal perturbations of ions and electrons can
be treated separately if
\begin{equation}
\omega \gg \frac{m_{e}}{m_{i}}\nu _{ei}.  \label{eq:7b}
\end{equation}
In this context it should be also noted that electron thermal effects in the
ionospheric E-layer are important for relatively low altitudes \citep{SN},
where ion magnetization is weak, whereas ion thermal effects become
important with strong ion magnetization.

\section{Farley-Buneman instability and ion-thermal effects}

Let as introduce dimensionless perturbations of electric potential, number
density, and temperature for the $\alpha $ species:
\begin{equation}
\bar{\phi}_{\alpha }=\frac{e\mathbf{k}\cdot \mathbf{E}^{\prime}}{\mathcal{K}
T_{\alpha }k^{2}},~~\bar{n}=\frac{n^{\prime }}{n},~~\bar{\tau}_{\alpha }=%
\frac{T_{\alpha }^{\prime }}{T_{\alpha }},  \label{eq:8}
\end{equation}
where primed variables stand for linear perturbations in the Fourier space,
and wave vector $k\perp b$ (here we considered only two dimensional
perturbations with wave vectors perpendicular to the background magnetic
field).

Then, linearizing Eqs. (\ref{eq:1})-(\ref{eq:3}), dropping viscosity and
thermal conductivity effects (these effects will be studied in the following
sections), setting for simplicity $\varepsilon =1$, and setting $%
T_{e}^{\prime }=0$, the Euler equation for the ions gives
\begin{equation}
\left( 1+\alpha ^{\ast }N-\frac{i\Omega }{\nu _{in}^{\ast }}\right) \mathbf{v%
}_{i}^{\prime }=-i\mathbf{k}\frac{u_{Ti}^{2}}{\nu _{in}^{\ast }}(\bar{\phi}%
_{i}+\bar{n}+\bar{\tau}_{i})+\kappa ^{\ast }\mathbf{v}_{i}^{\prime }\times
\mathbf{b}+\alpha ^{\ast }N\mathbf{v}_{e}^{\prime },  \label{eq:8a}
\end{equation}%
where $\Omega =\omega -\mathbf{k}\cdot \mathbf{V}_{i}$ is the frequency in
the ion frame, $\nu _{in}^{\ast }=\mu _{ni}\nu _{in}$ is the reduced
ion-neutral collisional frequency, $\kappa ^{\ast }=\kappa /\mu _{in}$, $%
\alpha ^{\ast }=\alpha /\mu _{in}$, and $u_{Ti}=(T_{i}/m_{i})^{1/2}$ is the
ion thermal velocity.

Similarly, the linearized Euler equation for the electrons (which we treat
inertialess) gives
\begin{equation}
\left( 1+N\right) \mathbf{v}_{e}^{\prime }=i\mathbf{k}\frac{u_{Ti}^{2}}{%
\alpha \nu _{in}^{\ast }}\left( \bar{\phi}_{i}-\frac{\gamma T_{e}}{T_{i}}%
\bar{n}\right) -\frac{\kappa }{\alpha }\mathbf{v}_{e}^{\prime }\times
\mathbf{b}+N\mathbf{v}_{i}^{\prime }.  \label{eq:8b}
\end{equation}

As discussed earlier, we study evolution of perturbations in the
limits $\alpha \ll 1$ and $\alpha N\ll 1$ that are
fulfilled in the entire chromosphere. In addition, here we assume also $\alpha N/\kappa ^{2}\ll 1$. This condition is valid everywhere
except for very low chromospheric heights, where the influence of Coulomb
collisions on FBI is negligible anyway \citep{GVPG09}. Solving equations (%
\ref{eq:8a})-(\ref{eq:8b}) for perturbed velocities and keeping only
leading-order terms with respect to the small parameters $\alpha $,
 $\alpha N$, and $\alpha N/\kappa ^{2}$, we obtain
\begin{equation}
\mathbf{v}_{i}^{\prime }=-i\left( \frac{u_{Ti}^{2}}{\nu _{in}^{\ast }}%
\right) \frac{(1-i\Omega /\nu _{in}^{\ast })\mathbf{k}+\kappa ^{\prime }%
\mathbf{k}\times \mathbf{b}}{(1-i\Omega /\nu _{in}^{\ast })^{2}+\kappa
^{^{\prime }2}}(\bar{\phi}_{i}+\bar{n}+\bar{\tau}_{i}),  \label{eq:9}
\end{equation}%
\begin{eqnarray}
\mathbf{v}_{e}^{\prime } =i\mathbf{k}\psi \frac{u_{Ti}^{2}}{\nu
_{in}^{\ast }}\left[ (1+N)\left( \bar{\phi}_{i}-\frac{\gamma T_{e}}{T_{i}}%
\bar{n}\right) \right.   \nonumber \\
\left. -\frac{N\kappa ^{\prime 2}(\bar{\phi}_{i}+\bar{n}+\bar{\tau}_{i})}{%
(1-i\Omega /\nu _{in}^{\ast })^{2}+\kappa ^{\ast 2}}\right] +\mathbf{Q}.
\label{eq:9a}
\end{eqnarray}%
Here $\mathbf{Q}$ stands for the terms proportional to $\mathbf{k}\times \mathbf{b}$, which do not contribute to the
dispersion relation (it is eliminated by the scalar product $\mathbf{k}%
\cdot \mathbf{v}_{e}^{\prime }$ in Eq. \ref{eq:13} below).

From equation (\ref{eq:9}) it follows that, in the lowest order with
respect to the small parameters $\alpha $, $\alpha N$, and
$\alpha N/\kappa ^{2}$, the perturbed ion velocity is not affected
by the Coulomb collisions with electrons. Physically, this means that the
force balance for the ion fluctuations is dominated by the ion-neutral
rather than the ion-electron collisions. In the leading order with respect
to small $\alpha $, $\alpha N$, and $\alpha N/\kappa ^{2}$, the perturbed equation (\ref{eq:3}) reduces to
\begin{equation}
\left( \frac{\Omega }{\nu _{in}^{\ast }}+i\zeta \right) \bar{\tau}_{i}-\frac{%
2\Omega }{3\nu _{in}^{\ast }}\bar{n}=\frac{4i\mu _{ni}}{3u_{Ti}^{2}}\mathbf{v%
}_{i}^{\prime }\cdot (\mathbf{V}_{i}-\mathbf{V}_{n}),  \label{eq:10}
\end{equation}%
where $\zeta =2\mu _{ni}+\nu _{ep}/\nu _{in}^{\ast }$.

From the ion continuity equation, using (\ref{eq:4e}) and (\ref{eq:9}), we
get
\begin{eqnarray}
\mathbf{v}_{i}^{\prime }\cdot (\mathbf{V}_{i}-\mathbf{V}_{n}) =\frac{%
\Omega \kappa ^{\prime }\bar{n}}{k^{2}(1-i\Omega /\nu _{in}^{\ast })}\times
\nonumber \\
\left[ (1-i\Omega /\nu _{in}^{\ast })\mathbf{k}\cdot (\mathbf{U}_{0}\times
\mathbf{b})+\kappa ^{\prime }(\mathbf{k}\cdot \mathbf{U}_{0})\right] .
\label{eq:11}
\end{eqnarray}%
Substituting this into (\ref{eq:10}) we obtain the relation between the
temperature and density perturbations $\bar{\tau}_{i}$ and $\bar{n}$:
\begin{eqnarray}
\frac{3kU_{0}}{2\Omega }\left( 1-i\frac{\Omega }{\nu _{in}^{\ast }}\right)
\left( \zeta -i\frac{\Omega }{\nu _{in}^{\ast }}\right) \bar{\tau}_{i} =%
\left[ \frac{2\mu _{ni}\kappa ^{2}U_{0}^{2}\cos \theta }{u_{Ti}^{2}}\right. +
\nonumber \\
\left. \left( 1-i\frac{\Omega }{\nu _{in}^{\ast }}\right) \left( \frac{%
2\mu _{ni}\kappa ^{2}U_{0}^{2}\sin \theta }{u_{Ti}^{2}}-i\frac{kU_{0}}{\nu
_{in}^{\ast }}\right) \right] \bar{n},  \label{eq:12}
\end{eqnarray}%
where $\theta $ is the angle between $\mathbf{U}_{0}$ and $\mathbf{k}$.

To obtain the second independent relation between $\bar{\tau}_{i}$ and $\bar{%
n}$, we use the electron continuity equation
\begin{equation}
(\Omega +\mathbf{k}\cdot \mathbf{U}_{0})\bar{n}=\mathbf{k}\cdot {\mathbf{v}%
_{e}^{\prime }}.  \label{eq:13}
\end{equation}%
Substituting (\ref{eq:9a}) in this equation gives
\begin{eqnarray}
-i\frac{k^{2}u_{Ti}^{2}}{\nu _{in}^{\ast 2}}\bar{\tau}_{i} =\left[ \frac{%
\Omega +\mathbf{k}\cdot \mathbf{U}_{0}}{\bar{\psi}\nu _{in}^{\ast }}+i\frac{%
k^{2}c_{s}^{2}}{\nu _{in}^{\ast 2}}+\right.   \nonumber \\
\left. \frac{\Omega }{\nu _{in}^{\ast }}\frac{(1-i\Omega /\nu _{in}^{\ast
})^{2}+\kappa ^{2}/(1+N)}{1-i\Omega /\nu _{in}^{\ast }}\right] \bar{n},
\label{eq:14}
\end{eqnarray}%
where $c_{s}=[\mathcal{K}(T_{i}+T_{e})/m_{i}]^{1/2}$ is the isothermal sound
speed and $\bar{\psi}=\psi (1+N)$. Note that in the absence of thermal
effects, $\bar{\tau}_{i}=0$, the expression in the square brackets on the
left hand side of Eq. (\ref{eq:14}) represents the dispersion relation for
isothermal electrostatic perturbations in weakly ionized plasmas studied by %
\citet{GVPG09}.

By means of the Eqs. (\ref{eq:12}) and (\ref{eq:14}), which represent two
independent relations between $\bar{\tau}_{i}$ and $\bar{n}$, one can
readily derive the dispersion relation. A simple analytical solution of the
dispersion equation can be obtained in the long-wavelength low-frequency
limit
\begin{equation}
|\Omega |,~kU_{0}\ll \nu _{in}^{\ast }.  \label{eq:15}
\end{equation}

\begin{figure}[tbp]
\centering
\includegraphics[width=5cm]{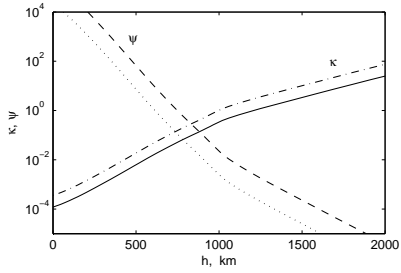}
\caption{Proton magnetization $\protect\kappa _{p}$ in the chromosphere for $%
B=30$ $\mathrm{G}$ (solid line) and $B=90$ $\mathrm{G}$ (dash-dot line), and
$\protect\psi $ for $B=30$ $\mathrm{G}$ (dashed line) and $B=90$ $\mathrm{G}$
(dotted line).}
\label{fig2}
\end{figure}

\begin{figure}[tbp]
\centering
\includegraphics[width=5cm]{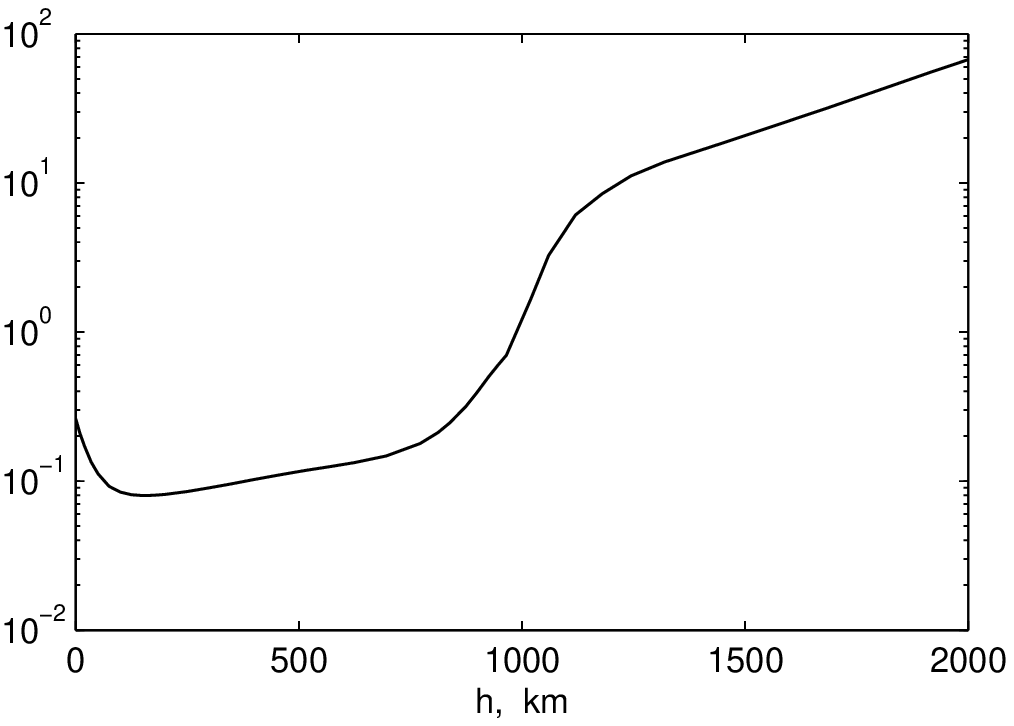}
\caption{The ratio $N$ of the Coulomb to electron-neutral collision
frequencies as a function of height.}
\label{fig1}
\end{figure}

\begin{figure*}[tbp]
\centering
\includegraphics[width=10cm]{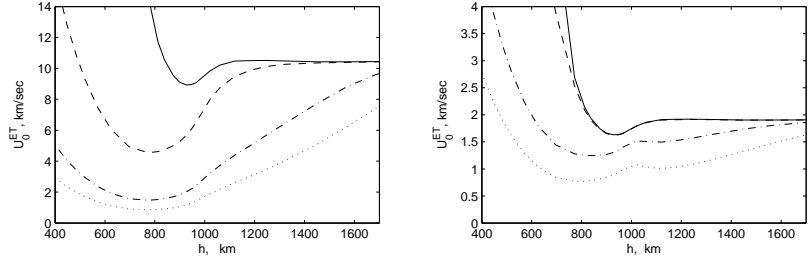}
\caption{Dependence of the FBI threshold $U_{0}^{ET}$ on the chromospheric
height for $\protect\varepsilon ^{\ast }=0$ (solid line), $\protect%
\varepsilon ^{\ast }=1$ (dashed line), $\protect\varepsilon ^{\ast }=10$
(dashed-dotted line) and $\protect\varepsilon ^{\ast }=30$ (dotted line).
Left panel corresponds to the protons and right panel to ions with $%
m_{i}=30m_{p}$. }
\label{fig3}
\end{figure*}

\begin{figure*}[tbp]
\centering
\includegraphics[width=10cm]{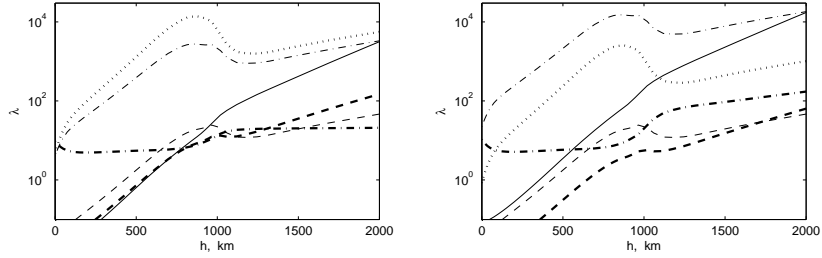}
\caption{The characteristic FBI wavelengths as functions of the
chromospheric height in the SRPM 306 model: $\protect\lambda _{n}$ (dotted
line), $\protect\lambda _{e}$ (thin dashed line), $\protect\lambda _{i}$
(thick dashed line), $\protect\lambda _{T}$ (thin dash-dotted line), $%
\protect\lambda _{\protect\kappa }$ (thick das-dotted line) and $\protect%
\lambda _{0}$ (solid line). Left panel corresponds to the protons and right
panel to ions with $m_{i}=30m_{p}$. }
\label{fig4}
\end{figure*}

Eliminating $\tau $ from Eqs. (\ref{eq:12}) and (\ref{eq:14}), and keeping
first-order terms in small parameters $|\Omega |/\nu _{in}^{\ast }$ and $%
kU_{0}/\nu _{in}^{\ast }$, we obtain the real part of frequency
\begin{equation}
\Omega _{r}=-\frac{\mathbf{k}\cdot \mathbf{U}_{0}}{1+\bar{\psi}}.
\label{eq:16}
\end{equation}%
Analysis of the second-order terms yields the following expression for the
growth rate
\begin{eqnarray}
\gamma  =\frac{\bar{\psi}k^{2}U_{0}^{2}}{\nu _{in}^{\ast }(1+\bar{\psi})}%
\left[ \frac{\left( 1-\frac{\kappa ^{2}}{1+N}\right) \cos ^{2}\theta }{(1+%
\bar{\psi})^{2}}-\frac{c_{s}^{2}}{U_{0}^{2}}+\right.   \nonumber \\
\left. \frac{4\mu _{ni}}{3\zeta }\frac{\kappa \cos \theta (\kappa \cos
\theta +\sin \theta )}{1+\bar{\psi}}\right].  \label{eq:17}
\end{eqnarray}%
Equations (\ref{eq:16})-(\ref{eq:17}) represent solution for the frequency
and the growth rate in the lowest order with respect to the small parameters
$\alpha $, $\alpha N$, and $\alpha N/\kappa ^{2}$.
Equations (\ref{eq:16})-(\ref{eq:17}) generalize Eqs. (29) and (30) from Dimant and Oppenheim (2004) by including Coulomb
collisions and allowing for different masses of the colliding ions and neutrals.

The first term in the square brackets of Eq. (\ref{eq:17}) drives the FBI,
while the last term drives the ion thermal instability.
If Coulomb collisions are ignored ($N=0$) then the driving term reduces to
the well known result by \citet{FPF84}, which implies that, regardless of
the neutral drag velocity, the FBI cannot occur if the proton magnetization $%
\kappa >1$. The dependence of the proton magnetization $\kappa $ on height
in the chromosphere based on the semi-empirical chromospheric model SRPM 306 %
\citep{FBH07} is shown in Fig. \ref{fig2} for $B=30\mathrm{G}$ (solid line)
and $B=90\mathrm{G}$ (dash-dot line), and $\psi $ for $B=30\mathrm{G}$
(dashed line) and $B=90\mathrm{G}$ (dotted line). It is seen that the proton
magnetization exceeds unity in the upper chromosphere and the standard FBI
theory predicts its stability there. In contrast, as was shown by %
\citet{GVPG09}, the Coulomb collisions make FBI possible even if ions are
relatively highly magnetized (the effect of reduced magnetization $\sim
\kappa ^{2}/(1+N)$ in the numerator). Detailed analysis shows that the
effect related to the Coulomb collisions makes the FBI possible for
chromospheric heights from $\sim 1000$ to $\sim 1400$ km (Fig. 2 by %
\citet{GVPG09}).

The dependence of $N=\nu _{ep}/\nu _{en}$ on the height for the model SRPM
306 is presented in Fig. \ref{fig1}. It is seen that Coulomb collisions
become dominant at heights $h>1000~\mathrm{km}$ and hence the development of
FBI is facilitated in the upper chromosphere \citep{GVPG09}.

If ion thermal effects are ignored, than the most unstable mode has $\theta
=0$ and the threshold value of the current velocity necessary to trigger the
FBI is given by
\begin{equation}
U_{0}^{cr}=c_{s}(1+\bar{\psi})\left( 1-\frac{\kappa ^{2}}{1+N}\right)
^{-1/2}.
\end{equation}%
Using the SRPM 306 model, \citet{GVPG09} found that the minimum value of $%
U_{0}^{cr}$ occurs at chromospheric height of $850~\mathrm{km}$ and is about
$2~\mathrm{km/s}$, which corresponds to the current $J_{0}\sim 2.4\times
10^{6}~\mathrm{statampere/cm^{2}}$. According to recent observations, the
typical values of currents at length scales $\sim 100~\mathrm{km}$ and
longer are much smaller, $\sim 5\times 10^{4}~\mathrm{statampere/cm^{2}}$ %
\citep{S07}. In principle it is possible that stronger currents exist
locally at smaller scales, but in this case the heat produced by the
ion-neutral friction will be at least one order of magnitude larger than the
energy required to sustain the radiative losses in the chromosphere.
Consequently, \citet{GVPG09} concluded the FBI can not be responsible for
chromospheric heating.

The ion thermal driving described by the last term in square brackets
of Eq. (\ref{eq:17}) becomes important for relatively high chromospheric altitudes
where the ion magnetization is strong. Analysis of Eq. (\ref{eq:17}) shows
that the most unstable mode propagates at the angle $\theta _{IT}$,
\begin{equation}
\tan \theta _{IT}=2\kappa (1+\bar{\psi})\bar{\mu}\left[ 3-\kappa ^{2}\left(
\frac{3}{1+N}-2\eta \right) \right] ^{-1}.  \label{eq:19}
\end{equation}%
Here
\begin{equation}
\bar{\mu}_{pi}=\frac{2\mu _{pi}}{\zeta }=\left( 1+\frac{m_{i}+m_{p}}{2m_{p}}%
\frac{\nu _{en}}{\nu _{in}}N\right) ^{-1}.  \label{eq:20}
\end{equation}%
For protons and heavy ions with $m_{i}=30m_{p}$
\begin{equation}
\bar{\mu}_{pp}=\frac{1}{1+4.59N},~~~\bar{\mu}_{pi}=\frac{1}{1+71.2N}.
\label{eq:21}
\end{equation}%
Using the data presented in Fig. 1 we conclude that both in the lower
chromosphere (where positive charges are dominated by heavy ions), and in
the upper chromosphere, the Coulomb collisions strongly reduce the ion
thermal effects and make them negligible in the chromospheric conditions.

\section{Electron thermal effects}

As is mentioned above, the electron thermal effects are important at
relatively low altitudes, where ion magnetization is still weak. Therefore
we treat the ions as unmagnetized, whereas the electrons are assumed to be
strongly magnetized, in which case $\mathbf{V}_{i}\approx \mathbf{U}%
_{0}\approx \mathbf{V}_{n}$. Manipulations with the Euler equation for
electrons under the condition $\omega _{ce}\gg \nu _{en}$ yield
\begin{equation}
\mathbf{v}_{e}^{\prime }=\frac{\omega \bar{n}}{k^{2}}\left( \mathbf{k}-\frac{%
\omega _{ce}}{\nu _{en}(1+N)+{\eta }_{e}k^{2}}\mathbf{k}\times \mathbf{b}%
\right) ,  \label{eq:22}
\end{equation}%
From the Euler equation for ions and from the continuity equation, dropping
the terms of order $\psi \kappa ^{2}\sim 2.6\times 10^{-3}$ we have
\begin{equation}
\mathbf{v}_{i}^{\prime }=-i\frac{\mathbf{k}u_{Ti}^{2}}{\nu _{in}^{\ast
}(1+\xi -i\Omega /\nu _{in}^{\ast })}\left( \bar{n}+\frac{T_{e}}{T_{i}}\bar{%
\phi}_{e}\right) =\frac{\mathbf{k}}{k^{2}}\Omega \bar{n},  \label{eq:23}
\end{equation}%
where $\xi =k^{2}{\eta }_{i}/\nu _{in}^{\ast }$. Substituting Eqs. (\ref%
{eq:22}) and (\ref{eq:23}) into the perturbed heat balance equation for
electrons, and using condition $\omega _{ce}\gg \nu _{en}$, we obtain
\begin{eqnarray}
\left( i-2\omega _{ce}\varepsilon \frac{(\mathbf{k}\times \mathbf{b})\cdot
\mathbf{U}_{0}}{{g}_{en}k^{2}u_{Te}^{2}}-\varepsilon \frac{2\nu _{en}N%
\mathbf{k}\cdot \mathbf{U}_{0}}{\omega k^{2}u_{Te}^{2}}\right) \bar{n}=
\nonumber \\
\left[ \frac{3}{2}i-\frac{\chi _{e}k^{2}}{n\omega }-\frac{3m_{e}\nu_{en}}{%
m_{p}\omega }\left( 1+\rho_{en}\right) \left( 1+\frac{m_{p}}{m_{i}}N\right) %
\right] \bar{\tau}_{e},  \label{eq:24}
\end{eqnarray}%
where ${g}_{en}=1+{\eta }_{e}k^{2}/\nu _{en}(1+N)$.

The second equation relating $\bar{n}$ and $\bar{\tau}_{e}$ can be obtained
by eliminating $\mathbf{v}_{e}$ and $\bar{\phi}_{e}$ from the Euler equation
for ions by means of Eqs. (\ref{eq:22}) and (\ref{eq:23}). This yields
\begin{eqnarray}
\bar{\tau}_{e} =-\bar{n}\frac{T_{i}}{T_{e}}\left[ \frac{c_{s}^{2}}{%
u_{Ti}^{2}}-i\frac{\nu _{in}^{\ast }}{k^{2}u_{Ti}^{2}}\times \right.
\nonumber \\
\left. \left\{ (1+\xi -i\Omega /\nu _{in}^{\ast })\Omega +\frac{\omega }{%
\bar{\psi}{g}_{en}}\right\} \right] .  \label{eq:25}
\end{eqnarray}%
Substitution of $\tau $ from Eq. (\ref{eq:25}) into Eq. (\ref{eq:24}) gives
the dispersion equation. As in the case of the ion thermal instability, we
consider only the relatively long-wavelength/low-frequency limit when $%
|\Omega |,~kU_{0},~\eta k^{2},~\xi k^{2}/n\ll \nu _{in}^{\ast }$. In this
limit we have the real part of frequency
\begin{equation}
\Omega _{r}=-\frac{\mathbf{k}\cdot \mathbf{U}_{0}}{1+(1+\xi )\bar{\psi}{g}_{en}}.  \label{eq:26}
\end{equation}%
Accounting for the terms that are second-order in $|\Omega |/\nu _{in}^{\ast
}$ yields the following expression for the growth rate
\begin{eqnarray}
\gamma  =\frac{{g}_{en}\bar{\psi}}{\nu _{in}^{\ast }[1+(1+\xi )\bar{\psi}{g%
}_{en}]}\left[ \Omega _{r}^{2}-k^{2}c_{s}^{2}+\right.   \nonumber \\
\left. \varepsilon \frac{m_{p}\nu _{pn}}{m_{i}\omega _{cp}}\frac{1+N}{%
\frac{m_{p}\chi k^{2}}{m_{e}n\nu_{en}}+3\left( 1+\rho _{en}\right) \left( 1+%
\frac{m_{p}}{m_{i}}N\right) }\frac{k^{2}U_{0}^{2}\sin 2\theta }{1+(1+\xi )%
\bar{\psi}{g}_{en}}\right].  \label{eq:27}
\end{eqnarray}

If the thermal conduction and viscosity effects can be ignored (conditions
for this assumption as well as analysis of other characteristic length
scales in the chromosphere are presented in the next section), than Eqs. (%
\ref{eq:26}) and (\ref{eq:27}) reduce to
\begin{equation}
\Omega _{r}=-\frac{\mathbf{k}\cdot \mathbf{U}_{0}}{1+\bar{\psi}};
\label{eq:28}
\end{equation}%
\begin{eqnarray}
\gamma =\frac{\bar{\psi}}{\nu _{in}^{\ast }(1+\bar{\psi})}\left[ \frac{%
k^{2}U_{0}^{2}\cos ^{2}\theta }{(1+\bar{\psi})^{2}}-\right.  \nonumber \\
\left. \varepsilon ^{\ast }\frac{(1+N)k^{2}U_{0}^{2}\sin 2\theta }{3\kappa
(m_{i}/m_{p}+N)(1+\bar{\psi})}-k^{2}c_{s}^{2}\right] ,  \label{eq:29}
\end{eqnarray}%
where the effective heating coefficient $\varepsilon ^{\ast }=\varepsilon
/\left( 1+\rho _{en}\right) $ represents the cumulative effect of two
counter-acting processes: wave heating/collisional cooling.

Note that in the lower chromosphere, dominated by heavy ions, electron
thermal effects are reduced (compared to the upper chromosphere) due to the
presence of the $m_{i}/m_{p}$ ratio in the denominator of the second term on
the right hand side of Eq. (\ref{eq:29}). Analysis of Eq. (\ref{eq:29})
shows that the propagation angle $\theta _{ET}$ for the most unstable mode
is given by
\begin{equation}
\tan 2\theta _{ET}=\frac{2}{3}\varepsilon ^{\ast }\frac{1+\bar{\psi}}{\kappa
}\frac{1+N}{m_{i}/m_{p}+N}.  \label{eq:30}
\end{equation}%
The threshold value of the current velocity is
\begin{equation}
U_{cr}^{ET}=c_{s}\sqrt{2}(1+\bar{\psi})\left[ 1+\sqrt{1+\left( \frac{2}{3}%
\varepsilon ^{\ast }\frac{1+\bar{\psi}}{\kappa }\frac{1+N}{m_{i}/m_{p}+N}%
\right) ^{2}}\right] ^{-1/2}.  \label{eq:30a}
\end{equation}

Dependence of the threshold value of the current velocity $U_{0}^{ET}$ on
height in the chromosphere based on SRPM 306 is shown in Fig. \ref{fig3} for
$\varepsilon ^{\ast }=0$ (this case corresponds to FBI in the conditions of
negligible ion magnetization), $1,~10,~30$. The magnetic field $B=30$ G. The
left panel corresponds to the protons and the right to the ions with $%
m_{i}=30m_{p}$.

From Fig. \ref{fig3} one can see that, in the case of protons, the electron
thermal effects cause a significant reduction of the threshold current
velocity even for $\varepsilon ^{\ast }=1$, when there is no any plasma
heating. For higher values of $\varepsilon ^{\ast }$ the reduction of the
threshold current velocity becomes very strong, and for $\varepsilon ^{\ast
}=30$ the threshold value of the cross-field current velocity decreases
about 10 times. However, our estimations, similar to those by \citet{GVPG09}%
, show that this threshold reduction is insufficient to make the FBI heating
comparable to the direct collisional heating by super-critical currents. It
must be also noted that $U_{cr}^{ET}$ is still much larger than the observed
chromospheric currents \citep{S07}.

In the case of heavy ions, the electron thermal effects are less important
and for $\varepsilon ^{\ast }=1$ the influence of electron thermal effects
on the FBI is negligible. But for higher values of $\varepsilon ^{\ast }$
the decrease in $U_{0}^{ET}$ becomes significant also in the case of heavy
ions.

\section{Typical length scales of the electrostatic instabilities in the
chromosphere}

In this section we study in detail the assumptions made in the analysis
presented above. We determine the typical length scales of the electrostatic
instabilities in the chromosphere. As mentioned in Sec. 2, perturbations of
the neutral component can be ignored under the condition (\ref{eq:7a}). The
equivalent condition for the perturbation wavelength is
\begin{equation}
\lambda \ll \lambda _{n}\equiv \frac{2\pi c_{s}}{\nu _{in}}\frac{n_{n}}{n}.
\label{eq:31}
\end{equation}%
The condition (\ref{eq:7b}) that ion and electron thermal perturbations can
be considered separately yields the condition for wavelength
\begin{equation}
\lambda \ll \lambda _{T}\equiv \frac{2\pi c_{s}}{\nu _{ep}}\frac{m_{i}}{m_{e}%
}.  \label{eq:32}
\end{equation}%
In the derivation of Eqs. (\ref{eq:28})-(\ref{eq:30a}) we ignored ion and
electron viscosity and electron thermal conductivity effects. From Eq. (\ref%
{eq:22}) it follows that electron viscosity effects can be ignored if $\nu
_{en}\gg \eta _{e}k^{2}$. Taking into account the expression for the
electron viscosity \citep{B65}
\begin{equation}
\eta _{e}=0.73\frac{\mathcal{K}T_{e}}{m_{e}\nu _{ep}},  \label{eq:33}
\end{equation}%
we find that the electron viscosity can be neglected under the following
condition
\begin{equation}
\lambda \gg \lambda _{e}\equiv 2\pi \left( \frac{1+N}{0.73}\frac{\nu
_{en}\nu _{ep}}{u_{Te}^{2}}\right) ^{-1/2}.  \label{eq:34}
\end{equation}%
According to Eq. (\ref{eq:23}), ion viscosity can be neglected if $\nu
_{in}^{\ast }\gg \eta _{i}k^{2}$. Noting that the ion viscosity (Braginskii
\cite{B65})
\begin{equation}
\eta _{i}=0.96\frac{\mathcal{K}T_{i}}{\sqrt{m_{e}m_{i}}\nu _{ep}},
\label{eq:35}
\end{equation}%
we conclude that the ion viscosity can be neglected if
\begin{equation}
\lambda \gg \lambda _{i}\equiv 2\pi \left( \frac{\nu _{pn}\nu _{ep}}{%
0.96u_{Tp}^{2}}\frac{m_{e}^{1/2}}{m_{i}^{1/2}}\right) ^{-1/2}.  \label{eq:36}
\end{equation}

The perpendicular heat conductivity of electrons is \citep{B65}
\begin{equation}
\chi _{e}=4.66\frac{n\mathcal{K}T_{e}\nu _{ep}}{m_{e}\omega _{ce}^{2}}.
\label{eq:37}
\end{equation}
Eq. (\ref{eq:24}) yields that the electron heat conductivity can be
neglected if
\begin{equation}
\lambda \gg \lambda _{\kappa }\equiv 2\pi \left[ 3\left( \frac{1}{N}+\frac{%
m_{p}}{m_{i}}\right) \frac{\omega _{ce}\omega _{cp}}{4.66u_{Te}^{2}}\right]
^{-1/2}.  \label{eq:38}
\end{equation}

Finally, the long wavelength approximation used to solve the dispersion
equation is valid when
\begin{equation}
\lambda \gg \lambda _{0}\equiv 2\pi \frac{c_{s}}{\nu _{in}^{\ast }}.
\label{eq:39}
\end{equation}

The characteristic wavelengths $\lambda _{n}$,~ $\lambda _{e}$, ~$\lambda
_{i}$,~$\lambda _{T}$,~$\lambda _{\kappa }$, and $\lambda _{0}$, as
functions of chromospheric height based on SRPM 306 are presented in Fig. %
\ref{fig4}. The left panel corresponds to protons and the right panel to
heavy ions with $m_{i}=30m_{p}$. The magnetic field $B=30$ G is assumed.
Transition from the lower chromosphere with the effective ion mass $%
m_{i}\sim 30m_{p}$ to the upper chromosphere with $m_{i}\sim m_{p}$ occurs
at the heights around 1000 km. This means the left panel of Fig. \ref{fig4}
shows correct scales at $h>$ 1000 km, and the right panel at $h<$ 1000 km.

Assuming that the super-critical currents can occur in the solar
chromosphere locally and generate FBI, from the right panel of Fig. \ref%
{fig4} we deduce that in the lower chromosphere, where the positively
charged particles are mainly heavy ions, the typical FBI wavelengths are $%
\lambda =10-10^{2}$ cm. In the upper chromosphere, where the positive charge
is dominated by protons, the characteristic wavelengths are $\lambda
=10^{2}-10^{3}$ cm (see left panel of Fig. \ref{fig4}). Since FBI generate
plasma density perturbations, they can generate plasma irregularities with
typical length scales $\sim 10-10^{2}$ cm in the lower, and $\sim
10^{2}-10^{3}$ cm in the upper chromosphere. These plasma irregularities
should cause scintillations of radio waves with similar wave lengths and
provide a tool for remote chromospheric sensing. In particular,
scintillations of decimetric/metric radio waves passing through solar
chromosphere can serve as indicators for FBI developed in lower/upper
chromosphere, and hence for the presence of over-threshold currents there.

\section{Discussion}

Since we interested in more general features of FB-type instabilities, we
did not analyze effects of inelastic electron-neutral collisions separately
but incorporated them into the effective heating parameter $\varepsilon
^{\ast }=\varepsilon /\left( 1+\rho _{en}\right) $. This parameter reflects
the response of electrons to the heating by waves ($\varepsilon $ in the
numerator) versus cooling by collisions ($1+\rho _{en}$ in the denominator).
Given the present uncertainty of both the heating factor $\varepsilon $ and
the inelastic collisional rates of electrons determining $\rho _{en}$ in the
chromosphere, the separate analysis of these effects is postponed for future
considerations. A more detailed and justified model is also needed
for the electron- and ion-neutral collisions in the chromospheric
conditions.

Several notes are in order regarding our study as compared to ionospheric
studies. We would like to emphasize here two important facts concerning
chromospheric plasma in contrast to ionospheric plasma: (i) Coulomb
collisions (represented by $N$) cannot be ignored in the chromosphere and
can increase the FBI growth rate; (ii) the ion/neutral mass ratio $%
m_{i}/m_{n}$ is large in the middle/lower chromosphere, which leads to the
decrease of the ion/neutral friction.

Since the Coulomb collisions usually introduce dissipative effects, their
favorable influence on FBI is counter-intuitive and needs some explanation.
As is known from ionospheric research \citep{OMO96,SN}, the destabilizing
term driving FBI is caused by the Pedersen response to the electric field
perturbations, whereas the stabilizing term (proportional to $\kappa
^{2}/(1+N)$) is related to the Hall response. The intervention of Coulomb
collisions in this picture is as follows: they abate the Pedersen term in
the growth rate less than the Hall term and thus facilitate the FBI making
it possible even for $\kappa >1$.

Without effects introduced by the Coulomb collisions and large ion/neutral
mass ratio (in the limit $N\rightarrow 0$ and $m_{i}/m_{n}\rightarrow 1$),
our results are compatible with the results of ionospheric E-layer research.
This conclusion follows from the comparison of our results on the thermal
FBI effects with results by \citet{DS95,DS97,R98,DO04}.

\section{Conclusions}

We investigated electrostatic instabilities of Farley-Buneman type in the
partially ionized plasma of the solar chromosphere taking into account ion
and electron thermal effects, electron and ion viscosity, and Coulomb
collisions. We derived the FBI growth rate including the ion thermal terms
and found that the Coulomb collisions highly reduce them in the middle/upper
chromosphere. Consequently, ion thermal effects can be neglected for FBI in
the solar chromosphere.

On the contrary, the electron thermal terms that contribute to the FBI
growth rate (\ref{eq:29}) are not negligible in the chromospheric conditions
and cause a significant reduction of the threshold current triggering the
instability. The ion and electron viscosity and thermal conductivity are
also important and reduce the instability growth rate for relatively
small-scale perturbations. We determined the characteristic length scales
relevant to chromospheric conditions well as the threshold value of the
current velocity as functions of height in the framework of the
semi-empirical chromospheric model SRPM 306.

it has to be noted that the study of \citet{GVPG09} did not take into account the effect of additional electron heating related to the presence of parallel electric field in waves. As showed theoreticaly by \citet{DM03} and confirmed by recent particle in cell simulations \citep{OD13}, this effect can significantly increase the electron heating. Importance of this mechanism for the solar chromosphere requires separate analysis and is out of the scope of this paper.

In spite of the considerable threshold reduction by the electron thermal
effects (see Eq. (\ref{eq:30a}) and Fig. \ref{fig3}),
our analysis showed that the electrostatic FB instabilities modified by the electron and ion thermal effects in
chromospheric conditions are less efficient heating mechanisms than the
collisional dissipation of cross-field currents that drive these
instabilities. This conclusion concerns both the lower chromosphere, where the threshold velocity is decreased by heavy ions, and the middle/upper chromosphere, where the threshold velocity is decreased
by the Coulomb collisions. As discussed in the introduction, our analysis ignored an additional electron heating related to the presence of parallel electric fields in waves. This effect is known to enhance significantly electron heating in the ionospheric E-layer and therefore we can not exclude the possibility that similar effect can take place in the solar chromosphere as well. This subject require further investigations. 

The characteristic wavelengths of the FB-type instabilities driven by
super-critical currents in the solar chromosphere are $\lambda =10-10^{3}$
cm. The plasma density fluctuations generated by these instabilities can
produce scintillations of radio waves propagating through the chromosphere.
The radio scintillations at $\sim 10$ cm wavelengths are indicators for the
FB instability developed in the lower chromosphere, while the scintillations
at $<10^{3}$ cm wavelengths suggest FBI in the upper chromosphere.
Observations and interpretations of such radio scintillations in terms of
FBI provide a possibility for remote diagnostics of strong cross-field
currents and plasma parameters in the solar chromosphere.

\section*{Acknowledgments}

This research was supported by the Belgian Federal Science Policy Office
(via IAP Programme - project P7/08 CHARM), by the European Commission's FP7
Program (projects 263340 SWIFF, 313038 STORM, and SOLAIRE Network
MTRN-CT-2006-035484), by FWO-Vlaanderen (project G.0304.07), and by
K.U.Leuven (projects C90347 and GOA/2009-009).

\end{document}